\documentclass[reprint, amsmath, amssymb, floatfix, aps, superscriptaddress, nofootinbib, nobibnotes, onecolumn]{revtex4-1}
\usepackage{amsfonts}
\usepackage{mathrsfs}
\usepackage{dblfnote}
\usepackage{graphicx}
\usepackage{hyperref}
\usepackage{bm}
\usepackage{amsmath}
\DeclareMathAlphabet{\mathscrbf}{OMS}{mdugm}{b}{n}
\usepackage{geometry}
\usepackage{indentfirst}
\usepackage{color}
\usepackage{acro}

\DeclareAcronym{LHAASO}{
    short = LHAASO,
    long = the Large High Altitude Air Shower observatory
}

\DeclareAcronym{HAWC}{
    short = HAWC,
    long = the High-Altitude Water Cherenkov Gamma-Ray Observatory
}

\begin{document}

\title{Pulsars as candidates of LHAASO sources J2226+6057, J1908+0621 and J1825-1326: \\ The leptonic origin}

\author{Zhe Chang} 
\email{changz@ihep.ac.cn}
\affiliation{Theoretical Physics Division, Institute of High Energy Physics, Chinese Academy of Sciences, Beijing 100049, China}
\affiliation{University of Chinese Academy of Sciences, Beijing 100049, China}

\author{Yu-Ting Kuang}
\email{kuangyt@ihep.ac.cn}    
\affiliation{Theoretical Physics Division, Institute of High Energy Physics, Chinese Academy of Sciences, Beijing 100049, China}
\affiliation{University of Chinese Academy of Sciences, Beijing 100049, China}

\author{Xukun Zhang}
\email{zhangxukun@ihep.ac.cn}    
\affiliation{Theoretical Physics Division, Institute of High Energy Physics, Chinese Academy of Sciences, Beijing 100049, China}
\affiliation{University of Chinese Academy of Sciences, Beijing 100049, China}

\author{Jing-Zhi Zhou} 
\email{zhoujingzhi@ihep.ac.cn}
\affiliation{Theoretical Physics Division, Institute of High Energy Physics, Chinese Academy of Sciences, Beijing 100049, China}
\affiliation{University of Chinese Academy of Sciences, Beijing 100049, China}

% 二阶诱导GW被广泛研究, 但未有人考虑一阶张量直接诱导二阶, 中微子 damping (未正确考虑!!!) 结论???? 得到了 全部解析形式, 并数值解, 结论???

\begin{abstract}
   Recently, from 12 $\gamma$-ray Galactic sources, the LHAASO has detected ultrahigh-energy photons up to 1.4PeV. 
The $\gamma$-ray spectra of the sources J2226+6057, J1908+0621, J1825-1326 and the suggested origin pulsars near the sources 
have been published. In our previous work, we studied the hadronic $\gamma$-ray spectra of the sources J2226+6057, J1908+0621, J1825-1326 
in terms of the Hertzian dipole model of pulsar. In this paper, we investigate the possibility of the leptonic origin of the 
$\gamma$-ray. We use the Hertzian dipole model to describe the pulsars around the sources. The electrons around the pulsars can be accelerated to PeV
by the electromagnetic fields of pulsars. Under the assumption that the initial electrons are uniform distributed in a
spherical shell between $10^{7}$ to $10^{9}$m around the pulsar,
we obtain the energy distribution of electrons. The leptonic $\gamma$-ray spectra can be calculated through inverse Compton scattering processes. {The leptonic $\gamma$-ray can roughly conform to the observation of LHAASO.}

\end{abstract}

\maketitle

\section{Introduction}\label{sec:intro}
% 1. 高能光子起源的相关研究  

  The origin of the high-energy $\gamma$-ray has been studied for a long time. There are two mainstream explanations. The first is the leptonic explanation: these $\gamma$-rays are produced by electrons through inverse Compton 
scattering (ICS). Another one is hadronic explanation: the protons will produce photons via $\pi$-decay. Both of those explanations 
can give an acceptable $\gamma$-ray spectrum under the assumption of certain initial leptonic/hadronic spectrum. However, why the leptonic/hadronic spectrum for injection 
should be chosen in such way still has no satisfied explanations. In addition, the acceleration mechanism of 
the leptonic/hadronic is still not clear. The shock wave produced by supernova explosion is the most popular 
acceleration mechanism \cite{Fermi:1949ee, Drury:1983zz, Schure:2012du}. Pulsar is another possible source for the high-energy $\gamma$-ray.
It is widely believed that there is a magnetosphere formed by plasmas around pulsar \cite{Goldreich:1969sb, Cheng:1986qt, Daugherty:1995zy, Dyks:2003rz, Muslimov:2004ig}. The magnetosphere structure has been 
studied through numerical simulation for many years \cite{Spitkovsky:2006np, Kalapotharakos:2008zc, Contopoulos:2009vm, Tchekhovskoy:2012hm, Philippov:2014mqa}.
In the Ref. \cite{Chang:2021bpc}, we studied the acceleration of charged particles around pulsars in terms of Hertzian magnetic dipole model of pulsars,
which can be used to study leptonic/hadronic $\gamma$-ray spectra. 

%2. 高能光子的观测。
The observation of high-energy photons is very difficult, since the survival rate of photons decays exponentially.
The main obstacle is caused by the $\gamma\gamma\to e^{-}e^{+}$ process, which has been fully studied for many years \cite{Gould:1967zza,DeAngelis:2013jna,Supanitsky:2013yja,DeAngelis:2016qob,Galanti:2019rnl,Dwek:2012nb}.	
Along with the development of technology, the PeV photons has been found finally. Photons around 0.1 PeV have been 
detected by the Tibet AS$\gamma$ \cite{Amenomori:2019rjd} and \ac{HAWC} \cite{Abeysekara:2021yum,HAWC:2019tcx}. Recently 
the Large High Altitude Air Shower observatory(LHAASO) has observed ultrahigh-energy photons even up to 1.4 PeV from 12 $\gamma$-ray 
Galactic sources \cite{cao_ultrahigh-energy_2021}. In Ref. \cite{Chang:2022fvj}, we showed that 
using the Hertzian magnetic dipole model of pulsars can give the hadronic $\gamma$-ray spectra, {which can fit with the observations of LHAASO.}
In this 
paper, we use the same acceleration mechanism to get the energy distributions of leptons and calculated the corresponding leptonic $\gamma$-ray spectra.
{We find that most of the leptonic $\gamma$-ray spectra of the suggested origin pulsars
can better meet the LHAASO observed $\gamma$-ray spectra of the sources J2226+6057,
J1908+0621, J1825-1326 than the hadronic situation.}
	
%3.扩散相关模型
	Focusing on the leptonic origin, the diffusion process will be very important, while this 
process can be ignored in the case of hadrons \cite{cao_ultrahigh-energy_2021}. Lots of models have been used to describe the diffusion process, such as 
slow-diffusion model, two-zone diffusion model \cite{Fang:2018qco, Hooper:2017gtd} and superdiffusion model \cite{perri2016transport, Volkov_2015}. 
In this paper, we choose the simplest model (slow-diffusion model), which will be described in detail in the Sec.\ref{sec:energy_spectrum}. 

The remaining part of this paper is organized as follows. In Sec.\ref{sec:acc}, we introduce the model of pulsar and acceleration process of electrons, where we 
show that our injected electronic spectra can be described by the function used in previous works.
In Sec.\ref{sec:energy_spectrum}, we describe our model of diffusion process and compare our result with the data of LHAASO. 
We summarize our main conclusions and give some discussions in Sec.\ref{sec:sum}.

\section{Acceleration process of electrons}\label{sec:acc}
% 1. 写脉冲星的加速模型  2. 写在本文用到的扩散方程以及扩散模型
  The first step is to obtain the energy distributions of the electrons.
In Ref. \cite{Chang:2021bpc}, we used the Hertzian magnetic dipole model to study the acceleration of charged particles around pulsars.
The exact solutions of electromagnetic fields around the pulsars are given by

\begin{equation}
\begin{aligned}
        \boldsymbol{B}(t, \boldsymbol{x})=&\frac{\mu_{0}}{4\pi}\left(-\frac{\boldsymbol{M}}{r^{3}}-\frac{1}{r^{2} c} \dot{\boldsymbol{M}}-\frac{1}{r c^{2}} \ddot{\boldsymbol{M}}+\frac{3 \boldsymbol{r}}{r^{5}}(\boldsymbol{r} \cdot \boldsymbol{M})\right. \\
        &\left.+\frac{3 \boldsymbol{r}}{r^{4}}\left(\boldsymbol{r} \cdot \frac{1}{c} \dot{\boldsymbol{M}}\right)+\frac{\boldsymbol{r}}{r^{3}}\left(\boldsymbol{r} \cdot \frac{1}{c^{2}} \ddot{\boldsymbol{M}}\right)\right)_{\mathrm{ret}}\ ,
\end{aligned}
\end{equation}
\begin{equation}
\begin{aligned}
    \begin{gathered}
        \boldsymbol{E}(t, \boldsymbol{x})=-\frac{\mu_{0}}{4\pi}\left(\dot{\boldsymbol{M}}+\frac{r}{c} \ddot{\boldsymbol{M}}\right)_{\mathrm{ret}} \times \frac{\boldsymbol{r}}{r^{3}}\ ,
    \end{gathered}
\end{aligned}
\end{equation}
where r is the distance from the pulsar, $\mathbf{M}$ is the magnetic dipole moment and lower corner mark ``ret'' means the time 
should be taken as $t-r/c$.
The charged particle in the electromagnetic field obeys the Landau-Lifshit (LL) equation \cite{1975Landau}
\begin{equation}\label{eq:ma}
  \begin{aligned}
    \begin{gathered}
      m a^{\mu}=e F^{\mu \nu} u_{\nu}+\tau_{0}\left(q \frac{\mathrm{d} F^{\mu \nu}}{\mathrm{~d} \tau} u_{\nu}+\frac{q^{2}}{m} P_{\nu}^{\mu} F^{\nu \alpha} F_{\alpha \beta} u^{\beta}\right),
    \end{gathered}
  \end{aligned}
\end{equation}
where
\begin{equation}
  \begin{aligned}
    \begin{gathered}
        P_{\nu}^{\mu}=\delta_{\nu}^{\mu}-\frac{u^{\mu} u_{\nu}}{c^{2}}\ ,\tau_{0}=\frac{2}{3} \frac{q^{2}}{4 \pi \epsilon_{0} m c^{3}}\ .
    \end{gathered}
  \end{aligned}
\end{equation}
 The $u_{\nu}$, m and q is the 4-velocity, mass and charge of the particle, respectively.
 In the first term of equation ~(\ref{eq:ma}), the $F^{\mu \nu}$ is the electromagnetic tensor. 
The second term of equation ~(\ref{eq:ma})
describes the influence of the particle’s own electromagnetic field. Assuming the electrons are uniform distributed 
from $10^{7}$ to $10^{9}$m around the center of pulsar, we can drive the final energy distribution of electrons, which will be used in 
diffusion process. 

\begin{figure}[htp]
  \centering
  \includegraphics[scale=0.5]{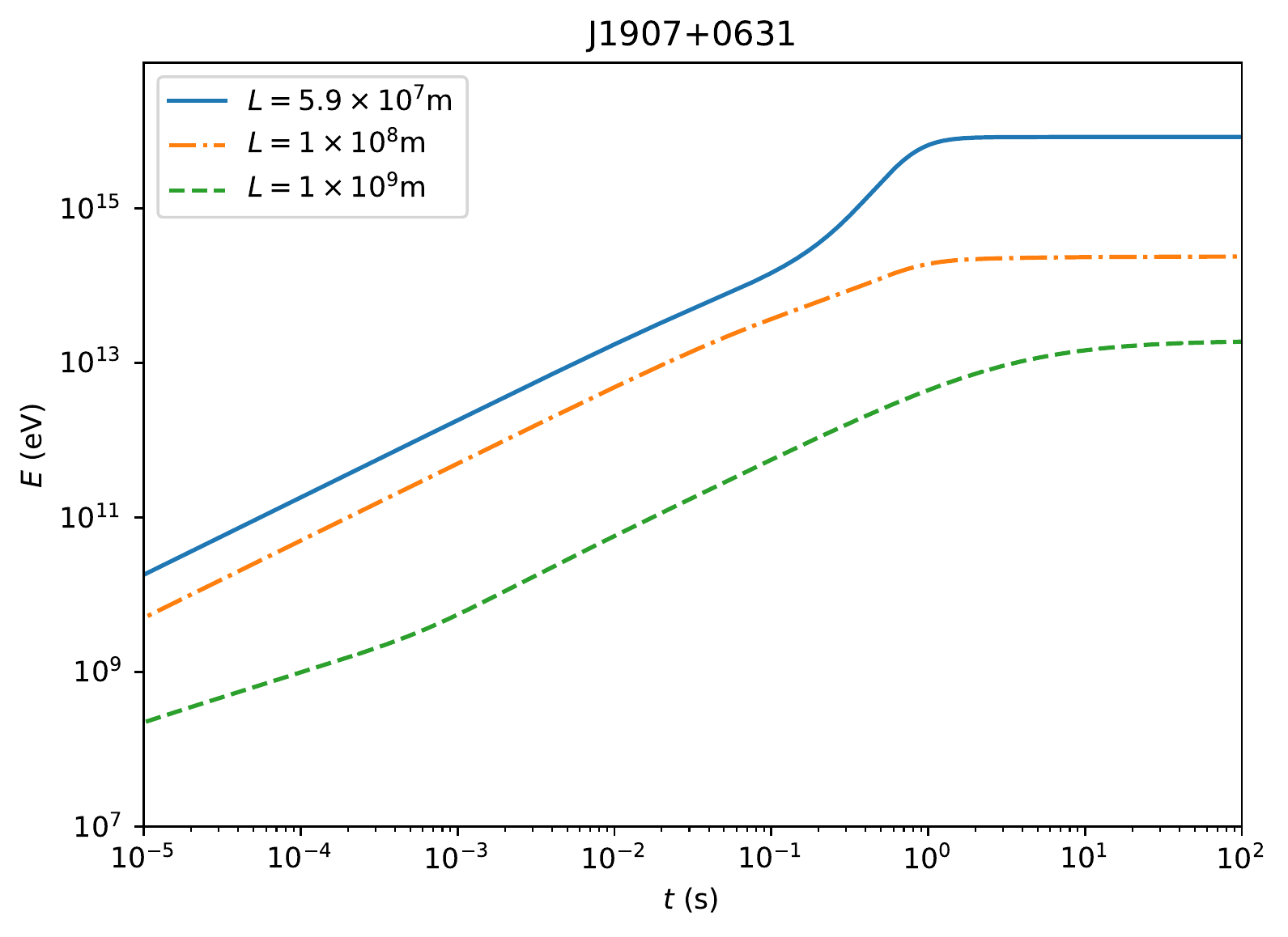}
  \caption{ The acceleration of electrons around a pulsar. We have set the inclination angle of pulsar to $\pi/6$, angular speed to $6.25\pi\ \mathrm{s}^{-1}$ 
  and the magnetic moment M to $2.42\times 10^{28}\mathrm{Am}^{2}$, as in the case of pulsar J1907+0631. Distance from each electron to the pulsar is marked in the upper left corner.}\label{fig:acc}
\end{figure}

In Fig.\ref{fig:acc}, we describe the acceleration of electrons at different initial position. We find that, in the Hertzian magnetic dipole model,
electrons can be accelerated to PeV within seconds. Besides, this result also shows the initial velocity have little influence
of the final energy of electrons, since their energy will remain almost constant after a few seconds.

\begin{figure}[htp]
  \centering
  \includegraphics[scale=0.4]{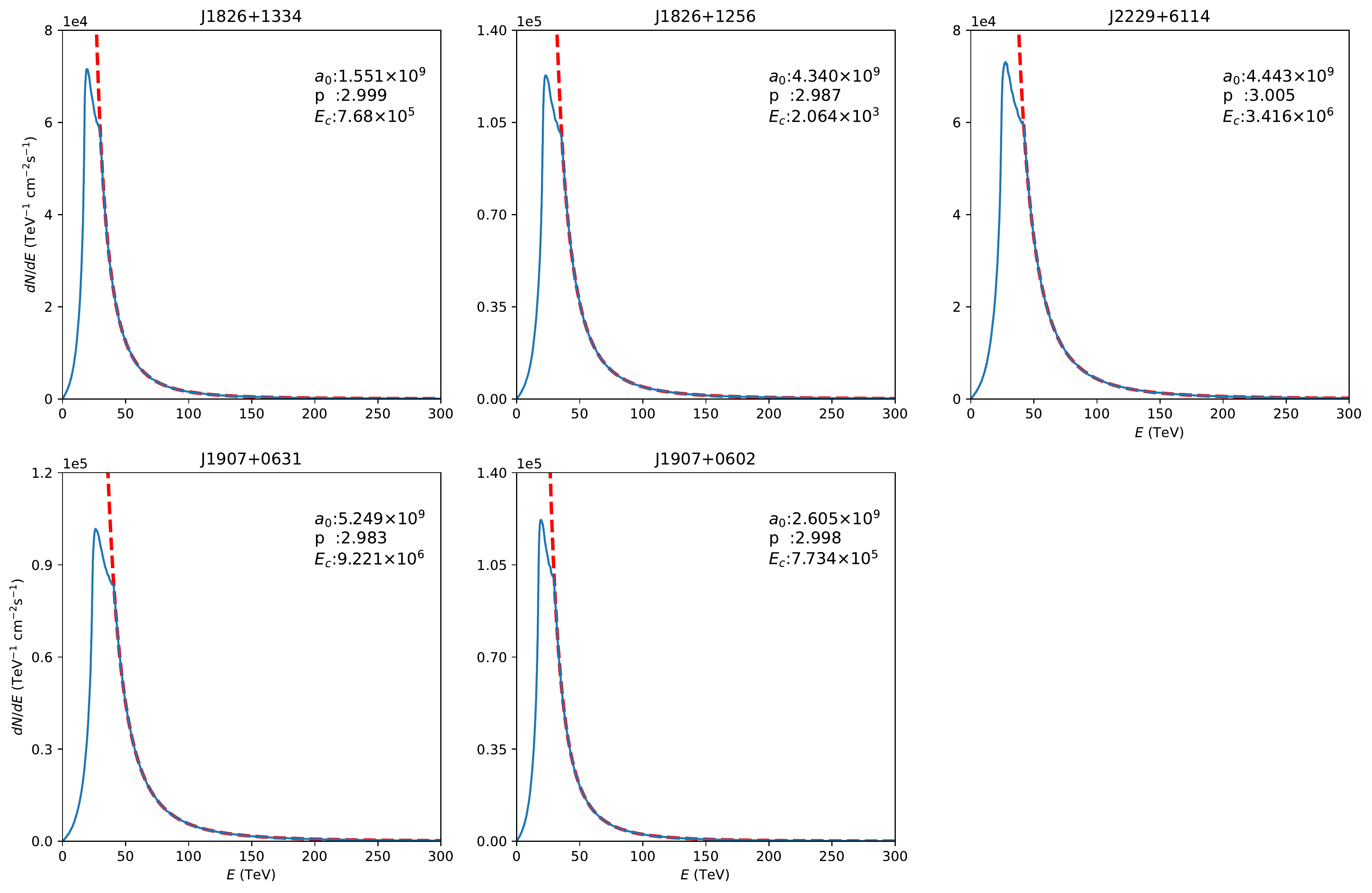}
  \caption{The solid curves are the electronic spectra from our simulation, while the dashed curves are described by $a_{0}E^{-p}e^{-(E/E_{c})^{2}}$.
  We have identified the parameters used in our simulation in the upper right corner and the $E_{c}$ we used is much greater than 
  previous hypothesis. In the high-energy part, this analyzing function can fit our results with adjusted $R^{2}$ bigger than 0.999.}\label{fig:fit}
\end{figure}

In Fig.\ref{fig:fit}, we show the energy distributions of electrons. We find that the high-energy 
parts of the electronic spectra can be described by the analytic expressions $a_{0}E^{-p}e^{-(E/E_{c})^{2}}$ or $a_{0}E^{-p}e^{-(E/E_{c})}$ with adjusted $R^{2}$ greater than 0.999, where $E_{c}$
is the cutoff energy. 
Both of those two expressions have been widely used to describe the injected electronic spectrum \cite{Breuhaus:2021vkr,cao_ultrahigh-energy_2021,Fang:2020dmi}. 
Besides, we find that the low-energy parts of energy distributions of electrons hardly affect the high-energy leptonic $\gamma$-ray spectra over 10 TeV.

\section{Diffusion process and the final \texorpdfstring{$\gamma$}\ -ray spectra}\label{sec:energy_spectrum}
For the diffusion process, the propagation equation of leptons is given by
\begin{equation}\label{eq:dN_dt}
\begin{aligned}
    \frac{\partial N(E_{e}, \boldsymbol{r}, t)}{\partial t}=&\nabla \cdot\left[D(E_{e}) \nabla N(E_{e}, \boldsymbol{r}, t)\right]\\
    &+\frac{\partial[b(E_{e}) N(E_{e}, \boldsymbol{r}, t)]}{\partial E_{e}}+Q\left(E_{e}, \boldsymbol{r}, t\right)\ ,
\end{aligned}
\end{equation}
where $N_{e}$ is the number of electrons per unit energy, $D(E_{e})$ is the
diffusion coefficient, and it takes the form of $D(E_{e})=D_{0}(E_{e}/100\ TeV)^{\delta}$, where $\delta$ is the energy index of the 
diffusion coefficient, and we set it to 1/3 as Kolmogorov’s theory \cite{Kolmogorov1991The,AMS:2016brs}. In this paper, we use the slow-diffusion model. 
In this model, we assume a space-independent $D_{0} = 3.2\times10^{27}\mathrm{cm}^{2}/\mathrm{s}^{-1}$. {We choose the same value of $D_0$ as in Ref.\cite{Fang:2022mdg}. The value of $D_0$ around the pulsars can be much smaller than its value in the interstellar medium (ISM) according to the observed spatial morphologies of the $\gamma$-ray emission \cite{HAWC:2017kbo,LHAASO:2021crt}. This phenomenon may be caused by the effects of turbulent scattering \cite{Malkov:2012qd}.} $b(E_{e})$ is the energy losing rate, and it is dominated by synchrotron and ICS. Relevant parameters can be found in 
Ref. \cite{Fang:2020dmi}. We take the magnetic field strength as 3$\mu G$ to calculate the synchrotron component \cite{1996ApJ...458..194M}. 
The final term $Q(E_{e}, \boldsymbol{r}, t)$ is the source term.

We choose source term as this form
\begin{equation}
\begin{aligned}
    Q\left(E_{e}, \boldsymbol{r}, t\right)=\left\{\begin{array}{ll}
        q\left(E_{e},t_{0}\right) \left[\left(t_{age}+t_{\mathrm{sd}}\right) /\left(t+t_{\mathrm{sd}}\right)\right]^{2}
        \\\times\delta\left(\boldsymbol{r}-\boldsymbol{r}_{p}\right), & t \geq 0 \\
        \\
        0,   &t<0
        \end{array}\right.\ ,
\end{aligned}
\end{equation}
where the $q(E_{e},t_{0})$ is the energy distribution of electrons we derived in Sec.\ref{sec:acc}. As the electrons are distributed at a distance of 
$10^{7}\sim 10^{9}$m from the pulsars, which is far smaller than the distance from those pulsars to Earth, we assume all the electrons are in the
same position. Thus, we use $\delta$ function here. The $t_{age}$ and $\mathbf{r}_{p}$ are the age and position of pulsar. We assume 
the injected electrons are  proportional to $[(t_{age}+t_{sd})/(t+t_{sd})]^{2}$, which is proportional to the spin-down 
luminosity of the pulsar and $t_{sd}$ is the pulsar spin-down time scale. In our research, the pulsar is quite young and 
the exact age of them are unknown, so we take $t_{age} = t_{sd}$.

 Using Green function method, we can find the solution to equation ~(\ref{eq:dN_dt}):
\begin{equation}
\begin{aligned}
    N\left(E_{e}, \boldsymbol{r}, t\right)=&\int_{R^{3}} d^{3} \boldsymbol{r}_{0} \int_{t_{\mathrm{ini}}}^{t} d t_{0} \frac{b\left(E_{e}^{\star}\right)}{b\left(E_{e}\right)} \frac{1}{\left(\pi \lambda^{2}\right)^{3 / 2}} \\
    &\times\exp \left[-\frac{\left(\boldsymbol{r}-\boldsymbol{r}_{0}\right)^{2}}{\lambda^{2}}\right] Q\left(E_{e}^{\star}, \boldsymbol{r}_{0}, t_{0}\right)\ ,
\end{aligned}
\end{equation} 
where
\begin{eqnarray}
    &E_{e}^{\star} \approx \frac{E_{e}}{\left[1-b_{0} E_{e}\left(t-t_{0}\right)\right]},\\
    &\lambda^{2}=4 \int_{E_{e}}^{E_{e}^{\star}} \frac{D\left(E_{e}^{\prime}\right)}{b\left(E_{e}^{\prime}\right)} d E_{e}^{\prime},\\
    &b_{0}=\frac{b(E_{e})}{E_{e}^2}\ .
\end{eqnarray}
Here t is the age of pulsar, $t_{ini} = max\{t-1/(b_{0}E_{e}),0\}$. We need the electron surface density 
$S_{e}(\theta)=\int_{0}^{\infty} N(l_{\theta})dl_{\theta}$ to calculate the photon spectra, which is integral over the 
direction of vision. We assume that photon field is isotropic and obeys the graybody distribution, then the ICS process of electrons 
can be described as fellow 
\begin{equation}
  \begin{aligned}
    \frac{d N_{\gamma}}{d \omega d t}=\frac{T^{3} m_{\mathrm{e}}^{3} c^{3} \kappa}{\pi^{2} \hbar^{3}} \int_{\epsilon_{\gamma} / T}^{\infty} \frac{d v_{\gamma}}{d \omega d N_{\mathrm{ph}} d t} \frac{x^{2} d x}{\mathrm{e}^{x}-1}\ ,
  \end{aligned}
\end{equation}
where $\omega$ is the energy of upscattered photon, $r_{0} = e^{2}/(m_{e}c^{2})$ is the electron classical radius. T is the 
temperature of the seed photon field, $\kappa$ is the dilution factor. $d v_{\gamma}/(d \omega d N_{ph}d t)$ is the scattering rate of IC. 
More details are in Ref. \cite{2014ApJ...783..100K}.

We use the $\texttt{Naima}$ code \cite{naima} to calculate the ICS process of $\gamma$-ray surface brightness $S_{\gamma}(\theta, E_{\gamma})$ 
in the direction of $\theta$ from $S_{e}(\theta)$. We choose the CMB as the seed photon field. We only assume CMB here is because
in the energy areas we care about, the cross-section for scattering higher-energy infrared and optical photons is strongly suppressed \cite{HAWC:2017kbo}.  
After integrating the $\gamma$-ray surface brightness over $\theta$: $\int_{0}^{\theta_{0}}S_{\gamma}(\theta, E_{\gamma})d\theta$, we can obtain the final leptonic $\gamma$-ray spectra, where $\theta_{0}$ is 
the intrinsic extent of the sources measured by LHAASO \cite{cao_ultrahigh-energy_2021}.

\begin{figure}[htp]
  \centering
  \includegraphics[scale=0.4]{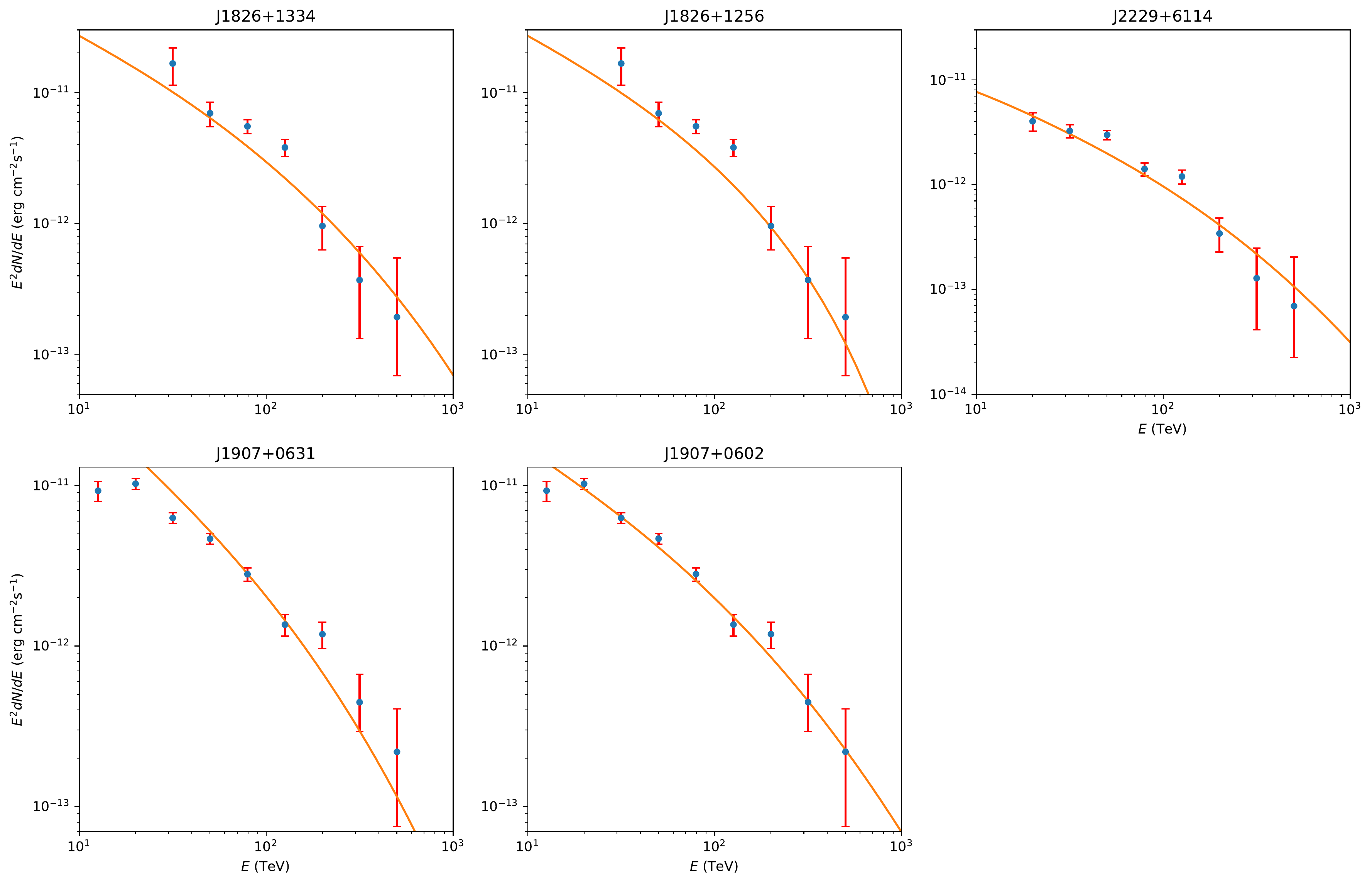}
  \caption{The origin curves are the leptonic $\gamma$-ray spectra derived from our model. The blue points are the data from LHAASO. 
  }\label{fig:fit_photon_spectrum}
\end{figure}

We simulate the acceleration of 5 million electrons and obtain the energy distributions of electrons.
We derive the leptonic $\gamma$-ray spectra  
of J2226+6057, J1908+0621, J1825-1326. For J1908+0621 and J1825-1326, there are two possible pulsars around each source,
we calculate separately for those pulsars. {We list some parameters of each pulsar in Table \ref{tab:parameters}}. The leptonic $\gamma$-ray spectra are shown in Fig.\ref{fig:fit_photon_spectrum}. 
{It shows that the leptonic $\gamma$-ray spectra of the suggested origin pulsars can be qualitatively consistent with the LHAASO observed $\gamma$-ray spectra of the sources J2226+6057,
J1908+0621, J1825-1326.}

\begin{table}[!ht]
  \caption{Some parameters of the five pulsars which are possible candidates of PeV $\gamma$-ray. Here, $P$ is the period of the pulsar, $\dot{P}$ is its first derivative. The surface magnetic field is noted as $B_s$, and the inclination angle is noted as $\theta$. These data can be found in Ref.\cite{Halpern:2001fc,Abdo:2010ht,Lyne:2016kam,Duvidovich:2019ykz,Fermi-LAT:2013svs}.}.
  \setlength{\tabcolsep}{12pt}
  \resizebox{14cm}{!}{ 
  \begin{tabular}{|c|c|c|c|} \hline
    PSR & $P(ms)$ & $\dot{P}(ss^{-1})$ & $B_s\sin\theta(G)$  \\ \hline
  J2229+6114 & 51.3 & $7.83\times 10^{-14}$ & $2.1\times 10^{12}$  \\ \hline
  J1907+0602 &	107 &	$8.68\times10^{-14}$ &	$3.0\times10^{12}$\\ \hline
  J1907+0631 &	324 &	$4.52\times10^{-13}$ &	$1.2\times10^{13}$\\ \hline
  J1826-1334 &	101 &	$7.52\times10^{-14}$ &	$2.8\times10^{12}$\\ \hline
  J1826-1256 &	110 &	$1.21\times10^{-13}$ &	$3.7\times10^{12}$\\ \hline
  \end{tabular}\label{tab:parameters}
  }
  
\end{table}

\section{CONCLUSION AND DISCUSSION}\label{sec:sum}
% 1. 做的事,  2. 磁层、引力忽略  2.1 磁层影响   3. 未来

In our model, we assumed that the electrons distributed uniformly in a spherical shell between $10^{7} \sim 10^{9}$m around the 
pulsar to avoid the impacts of possible magnetosphere structure. 
We studied the leptonic $\gamma$-ray spectra of the sources J2226+6057, J1908+0621, J1825-1326 in terms of the Hertzian magnetic 
dipole model of pulsars. Using the energy distributions of electrons we obtained, we considered the diffusion process in the slow-diffusion model 
and calculated the corresponding $\gamma$-ray spectra by using the $\texttt{Naima}$ code. {We concluded that the leptonic $\gamma$-ray spectra of the suggested origin pulsars could roughly describe the 
LHAASO observed $\gamma$-ray spectra of the sources J2226+6057, J1908+0621, J1825-1326.}
We also showed that our injection of electronic spectra in high-energy part consisted with the hypothesis of previous studies.

{In order to compare our results of leptonic spectra with the counterparts of hadronic spectra, we calculated the $\chi^2$ of each curve for both cases in Table \ref{tab:chi}. We find that the leptonic explanation is better than the hadronic explanation for the J1836-1334, J2229+6114 and J1907+0602. {For the J1826-1256, the leptonic spectra and hadronic spectra have the same level of fitting.} While, for the J1907+0631, the hadronic explanation has better fit. It is possible that only the hadronic explanation is applicable to this situation. This can be a meaningful topic in the future.}

\begin{table}[!ht]
  \caption{The $\chi^2$ for each curve in Fig.\ref{fig:fit_photon_spectrum} and the counterparts of the hadronic spectra in Ref. \cite{Chang:2022fvj}}.
  \resizebox{\textwidth}{!}{ 
  \begin{tabular}{|c|c|c|c|c|c|} \hline
    & J1826-1334 & J1826-1256 & J2229+6114 & J1907+0631 & J1907+0602  \\ \hline
  leptonic explanation & 16.87 & 20.92 & 19.12 & 225.88 & 19.73  \\ \hline
  hadronic explanation &	65.65 &	17.40 &	36.59 & 35.75 & 44.59\\ \hline
  \end{tabular}\label{tab:chi}
  }
  
\end{table}

\section*{Acknowledgement}
We thank Dr. Q.H. Zhu, Dr. K. Fang, and Prof. X.J. Bi for useful discussions. This work has been funded by the National Nature Science Foundation of China under grant No. 12075249 and 11690022, and the Key Research Program of the Chinese Academy of Sciences under Grant No. XDPB15.

\bibliography{biblio}

\end{document}